\documentclass[10pt,twocolumn,showpacs,longbibliography,amsmath,amssymb,osajnl,floatfix,superscriptaddress]{revtex4-1}

\usepackage{graphicx}
\usepackage{dcolumn}
\usepackage{bm}
\usepackage{color}
\usepackage{txfonts}
\usepackage{microtype}
\usepackage{ulem}

\begin{document}

\title{Stochasticity in radiative polarization of ultrarelativistic electrons in an ultrastrong laser pulse}

\author{Ren-Tong Guo}  \thanks{These authors have contributed equally to this work.}	\affiliation{MOE Key Laboratory for Nonequilibrium Synthesis and Modulation of Condensed Matter, School of Science, Xi'an Jiaotong University, Xi'an 710049, China}\affiliation{Max-Planck-Institut f\"{u}r Kernphysik, Saupfercheckweg 1,	69117 Heidelberg, Germany}
\author{Yu Wang}  \thanks{These authors have contributed equally to this work.}	\affiliation{MOE Key Laboratory for Nonequilibrium Synthesis and Modulation of Condensed Matter, School of Science, Xi'an Jiaotong University, Xi'an 710049, China}\affiliation{Max-Planck-Institut f\"{u}r Kernphysik, Saupfercheckweg 1,	69117 Heidelberg, Germany}
\author{Rashid Shaisultanov} \affiliation{Max-Planck-Institut f\"{u}r Kernphysik, Saupfercheckweg 1,
	69117 Heidelberg, Germany}
\author{Feng Wan}	\affiliation{MOE Key Laboratory for Nonequilibrium Synthesis and Modulation of Condensed Matter, School of Science, Xi'an Jiaotong University, Xi'an 710049, China}	
\author{Zhong-Feng Xu}	\affiliation{MOE Key Laboratory for Nonequilibrium Synthesis and Modulation of Condensed Matter, School of Science, Xi'an Jiaotong University, Xi'an 710049, China}
\author{Yue-Yue Chen}\email{yueyuechen@shnu.edu.cn}\affiliation{Department of Physics, Shanghai Normal University, Shanghai 200234, China}
\affiliation{Max-Planck-Institut f\"{u}r Kernphysik, Saupfercheckweg 1,
	69117 Heidelberg, Germany}		
\author{Karen Z. Hatsagortsyan}\email{k.hatsagortsyan@mpi-hd.mpg.de}
\affiliation{Max-Planck-Institut f\"{u}r Kernphysik, Saupfercheckweg 1,	69117 Heidelberg, Germany}

\author{Jian-Xing Li}\email{jianxing@xjtu.edu.cn}
\affiliation{MOE Key Laboratory for Nonequilibrium Synthesis and Modulation of Condensed Matter, School of Science, Xi'an Jiaotong University, Xi'an 710049, China}\affiliation{Max-Planck-Institut f\"{u}r Kernphysik, Saupfercheckweg 1,	69117 Heidelberg, Germany}

\date{\today}

\begin{abstract}

Stochasticity effects in the  spin (de)polarization of an ultrarelativistic electron beam during photon emissions in a counterpropoagating ultrastrong focused laser pulse in the quantum radiation reaction regime are investigated.
We employ a Monte Carlo method to describe the electron dynamics semiclassically,
and  photon emissions as well as the electron radiative polarization quantum mechanically.
While in the latter the  photon emission is inherently stochastic, we were able to identify its imprints  in comparison with the new developed semiclassical stochasticity-free method of radiative polarization applicable in the quantum regime. With an initially spin-polarized electron beam, the stochastic spin effects are seen in the dependence of the depolarization degree on the electron scattering angle and the electron final energy (spin stochastic diffusion). With an initially unpolarized electron beam,  the spin stochasticity is exhibited in enhancing the known effect of splitting of the electron beam along the propagation direction into two oppositely polarized parts by an elliptically polarized laser pulse. The considered stochasticity effects for the spin are observable  with   currently achievable laser and electron beam parameters.

\end{abstract}

\maketitle

The modern ultrastrong laser technique \cite{Danson__2019,Yoon2019,Gales_2018,ELI,Vulcan,Exawatt,CORELS} allows for investigation of nonlinear QED processes \cite{Bula_1996,Burke_1997,Turcu_2019} and radiation reaction effects \cite{Abraham_1905, Lorentz_1909, Dirac_1938, Heitler_1941,RMP_2012}. Thus, recently   classical and quantum signatures of radiation reaction in the electron energy losses have been identified in the experiments of an ultrarelativistic electron beam collision with strong laser fields \cite{Cole2018prx,Poder2018prx}.  Quantum features of radiation reaction originate from the discrete and probabilistic character of a photon emission, which gives rise to stochasticity effects. The latter is responsible for the broadening of the energy spread of an electron beam in a plane laser field  \cite{Piazza2013,Piazza2014,Yoffe_2015}, causes electron stochastic heating in a standing laser field \cite{Bashinov2015}, results in quantum quenching of radiation losses in short laser pulses \cite{Harvey2017}, disturbs the angular distribution of radiation \cite{Piazza2010,Li2017}, and brings about the so-called electron straggling effect \cite{Shen_1972,Blackburn_2014}, when the electron propagates a long distance without radiation, resulting in the increase of the yield of high-energy photons. Radiation reaction can have impact on the electron spin dynamics, proved since long ago for synchrotron radiation. It may induce polarization of the unpolarized electron beams (Sokolov-Ternov effect) \cite{Sokolov_1964,Sokolov_1968,Baier_1967,Baier_1972}, or depolarization of the initially polarized beam \cite{Derbenev_1972,Heinemann_2001}.

Polarized electrons are commonly generated either by accelerating nonrelativistic polarized electrons, obtained from  photocathodes \cite{Pierce_1976}, spin filters \cite{Batelann_1999} and beam splitters \cite{Dellweg_2017}, by conventional accelerators \cite{Swartz_1988} and laser wakefield accelerators \cite{Wen_2019, Wu_2019}, or by radiative polarization  in storage rings \cite{Derbenev_1973,Mane_1987a}, in which the polarization  typically requires  a period  from minutes to hours because of the modest magnetic fields in  storage rings $\sim$Tesla.
Recently, there are several proposals on how to use ultrastrong laser fields for generation of polarized relativistic electron beams \cite{Sorbo_2017,Sorbo_2018,Seipt_2018,li2019prl,Wan_2019plb,Li_2019spin,Chen_2019, Song_2019, Seipt_2019}. In particular,   the possibilities for creation of ultrarelativistic high-polarization high-density electron and positron beams  in femtoseconds via utilizing asymmetric spin-dependent radiation reaction in elliptically polarized laser fields are shown in \cite{li2019prl,Wan_2019plb,Li_2019spin}, and using two-color laser fields in  \cite{Chen_2019, Song_2019, Seipt_2019}.  The methods put forward in these works open a door for detailed investigation of all features of the radiative polarization and depolarization processes in ultrastrong focused laser fields,  as well as in multiple laser beam configurations. Usually, a full quantum mechanical study of radiation reaction includes all quantum effects, such as the photon recoil, stochasticity, and interferences, which makes difficult to single out the specific radiation reaction signatures of the stochasticity.

\begin{figure}[t]
  	\setlength{\abovecaptionskip}{-0.0cm}
 	\includegraphics[width=1.0\linewidth]{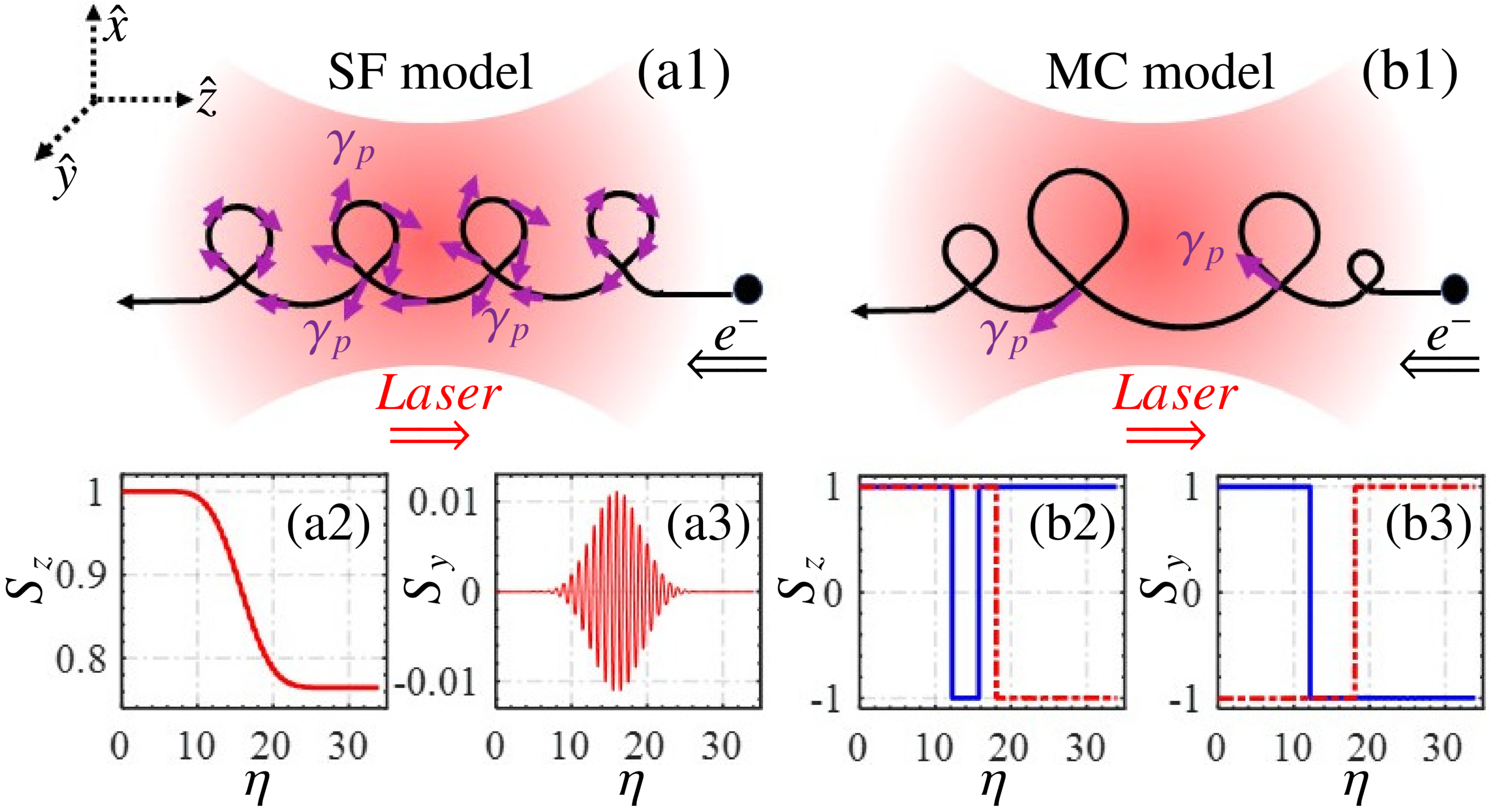}
 \caption{ Scenario for the detection of  stochasticity effects in radiative depolarization. The trajectories and spin evolutions of  LSP  electrons head-on colliding with a linearly polarized laser pulse, polarizing in $x$ direction and propagating along $+z$ direction: (a1)-(a3) in the SF model and (b1)-(b3) in the MC model. $S_\alpha$ is the spin component measured along the $\alpha$-axis, $\alpha=\{x,y,z\}$,
$e^-$ and $\gamma_p$ indicate the electron and emitted photon, respectively, red-dashed and blue-solid curves in (b2) and (b3)  two different sample electron, respectively, and $\eta$ the laser phase.
} \label{fig1}
 \end{figure}

In this work, stochasticity effects in radiative (de)polarization  of an ultrarelativistic electron beam head-on colliding with an ultrastrong laser pulse are investigated in the quantum radiation reaction regime;
see the interaction scenarios in Fig.~\ref{fig1}. We employ a Monte Carlo (MC) method to describe spin-resolved electron dynamics in a strong laser field, stochastic photon emissions, and corresponding stochastic radiative spin evolution.
To elucidate the role of stochastic spin effects, we develop an auxiliary semiclassical stochasticity-free (SF) method for the description of the spin-dependent radiation reaction on the electron dynamics. For this purposes we use the Baier-Katkov-Strakhovenko equation for  the expectation value of the electron spin \cite{Baier_1970,Baier_1972}, which is a generalization of the Thomas-Bargmann-Michel-Telegdi (TBMT) equation \cite{Thomas_1926, Thomas_1927, Bargmann_1959}, including radiation reaction for the electron spin. The latter is supplemented with the modified Landau-Lifshitz equation \cite{li2019prl}, including the spin-dependent radiation reaction and the quantum recoil.
Firstly, we consider a depolarization scenario for the initially longitudinally spin-polarized (LSP) (along velocity, $z$-axis) electron beam. The depolarization proceeds in different ways in semiclassical and quantum models, which after the interaction yields differences in the angle-resolved polarization distribution of the electron beam, and provides the stochasticity signatures for the spin radiative dynamics. In particular,  in the SF model an electron continuously loses energies due to radiation (without stochastic effects and
straggling in photon emissions, the photon energies are typically
low), which gradually alter the electron trajectory and the spin longitudinal component due to radiation reaction, while the spin component along the laser magnetic field
oscillates in the symmetric laser field,
as shown in Figs.~\ref{fig1}(a1)-(a3).
On the contrary, in the MC model
a finite number of photons are stochastically emitted with random energies and discretely alter the electron dynamics due to the quantum recoil,  and the quantum spin state stochastically flips on the instantaneous  spin quantization axis (SQA), see  Figs.~\ref{fig1}(b1)-(b3). The signatures of the stochasticity are identified by analyzing the features of the stochastic spin diffusion.
In the second applied scenario, we use an initially unpolarized electron beam head-on colliding with an elliptically polarized laser pulse. Here, the unpolarized electron beam splits along the propagation direction into two oppositely polarized parts, and the stochasticity effect is observed in enhancing the separation.

We employ ultrastrong laser fields with the invariant  field parameter $a_0\equiv eE_0/(m\omega_0)\gg1$ \cite{Ritus_1985,Baier1998}, where  $E_0$ and $\omega_0$ are the laser field amplitude and frequency, respectively, and $-e$ and $m$ the electron charge and mass, respectively. Relativistic units with $c=\hbar=1$ are used throughout.
The quantum radiation reaction regime requires the invariant quantum parameter $\chi\equiv |e|\sqrt{-(F_{\mu\nu}p^{\nu})^2}/m^3\gtrsim 1$ \cite{Ritus_1985, Baier1998},
with the field tensor $F_{\mu\nu}$ and the four-vector of the electron momentum $p^{\nu}$. In the electron-laser counterpropagating scheme,  $\chi\approx 2a_0\gamma_e\omega_0/m$, with the electron Lorentz factor $\gamma_e$.

In the MC method, we treat spin-resolved electron dynamics semiclassically and photon emissions quantum mechanically in the local constant field approximation \cite{Ritus_1985,Baier1998,Ilderton2019prd, piazza2019},  valid at $a_0\gg 1$.
At each simulation step, the photon emission is  calculated following the common algorithms~\cite{Ridgers_2014,Elkina2011,Green2015} and  the photon polarization, represented by the Stokes parameters  ($\xi_1$, $\xi_2$, $\xi_3$) \cite{McMaster_1961}, is calculated according to our MC algorithm introduced in Refs.~\cite{Ligammaray_2019, Wan_2020}; see also \cite{supplemental}.
After the photon emission the electron spin state is determined by the spin-resolved emission probabilities, derived in the QED operator method of Baier-Katkov \cite{Baier_1973},
and  instantaneously  collapsed into one of its basis states defined with respect to the instantaneous SQA, which is chosen according to the particular observable of interest: to determine the polarization of the electron along the magnetic field in its rest frame, the SQA is chosen along  the magnetic field ${\bm n}_B={\bm\beta}\times\hat{{\bf a}}$ with the scaled electron  velocity ${\bm \beta}={\bf v}/c$ and the unit vector $\hat{{\bf a}}={\bf a}/|{\bf a}|$ along the electron acceleration  ${\bf a}$
\cite{li2019prl,Chen_2019}. In the case when the electron beam is initially  polarized with the initial spin vector ${\bf S}_i$,  the observable of interest is the spin expectation value along the initial polarization and the SQA is chosen along that direction \cite{supplemental}.
 Between photon emissions, the spin precession is governed by TBMT equation:
\begin{eqnarray}\label{spin}
\left(\frac{{\rm d}{\bf S}}{{\rm d}\eta}\right)_{T}&=&\frac{e\gamma_e}{\left(k\cdot p_i\right)}{\bf S}\times\left[-\left(\frac{g}{2}-1\right)\frac{\gamma_e}{\gamma_e+1}\left({\bm \beta}\cdot{\bf B}\right)\cdot{\bm \beta}\right.\nonumber\\
&&\left.+\left(\frac{g}{2}-1+\frac{1}{\gamma_e}\right){\bf B}-\left(\frac{g}{2}-\frac{\gamma_e}{\gamma_e+1}\right){\bm \beta}\times{\bf E}\right],
\end{eqnarray}
where ${\bf E}$ and  ${\bf B}$ are the  laser electric and magnetic fields, respectively,  $\eta=k\cdot r$ the laser phase, $p_i$, $k$, and $r$   4-vectors of the electron momentum  before radiation, laser wave vector, and coordinate, respectively,  and
$g$ the electron gyromagnetic factor \cite{supplemental}.  The simulation results of the electron spin dynamics with our method concur with those of the CAIN code~\cite{CAIN}.

In our SF method, we revise the TBMT equation, including a term responsible for radiation reaction. For the revision we generalize for arbitrary $\chi$ the method of Refs.~\cite{Baier_1970,Baier_1972}, where radiation reaction for the spin evolution is calculated at $\chi\ll 1$.
Thus, the equation which is used for the spin evolution including radiation reaction reads:
\begin{eqnarray}\label{spin2}
\frac{{\rm d}{\bf S}}{{\rm d}\eta}&=&\left(\frac{{\rm d}{\bf S}}{{\rm d}\eta}\right)_{T}-P\left[\psi_1(\chi){\bf S}+\psi_2(\chi)({\bf S}\cdot{\bm \beta}){\bm \beta}+\psi_3(\chi){\bm n}_B\right],
\end{eqnarray}
where $P=\alpha m^2/[\sqrt{3}\pi\left(k\cdot p_i\right)]$, $\psi_1(\chi)$ = $\int_{0}^{\infty} u''{\rm d}u {\rm K}_{\frac{2}{3}}(u')$, $\psi_2(\chi)$ = $\int_{0}^{\infty} u''{\rm d}u \int_{u'}^{\infty}{\rm d}x{\rm K}_{\frac{1}{3}}(x)$-$\psi_1(\chi)$, $\psi_3(\chi)$ = $\int_{0}^{\infty} u''{\rm d}u {\rm K}_{\frac{1}{3}}(u')$,  $u'=2u/3\chi$, $u''=u^2/(1+u)^3$, $u=\varepsilon_\gamma/\left(\varepsilon_i-\varepsilon_\gamma\right)$,  $\varepsilon_i$ and $\varepsilon_\gamma$ are the electron energy before radiation and emitted photon energy, respectively, and ${\rm K}_n$  the $n$-order modified Bessel function of the second kind. Further, in the SF method
the electron dynamics is described by the Landau-Lifshitz equation \cite{Landau1975} with corrections for the quantum recoil \cite{Piazza2012}, and the photon polarization is calculated by the average method as in \cite{Ligammaray_2019}.  The validity, comparison and more details on  MC and SF methods are given in \cite{supplemental}.

\begin{figure}[t]
	\setlength{\abovecaptionskip}{-0.0cm}
	\includegraphics[width=1.0\linewidth]{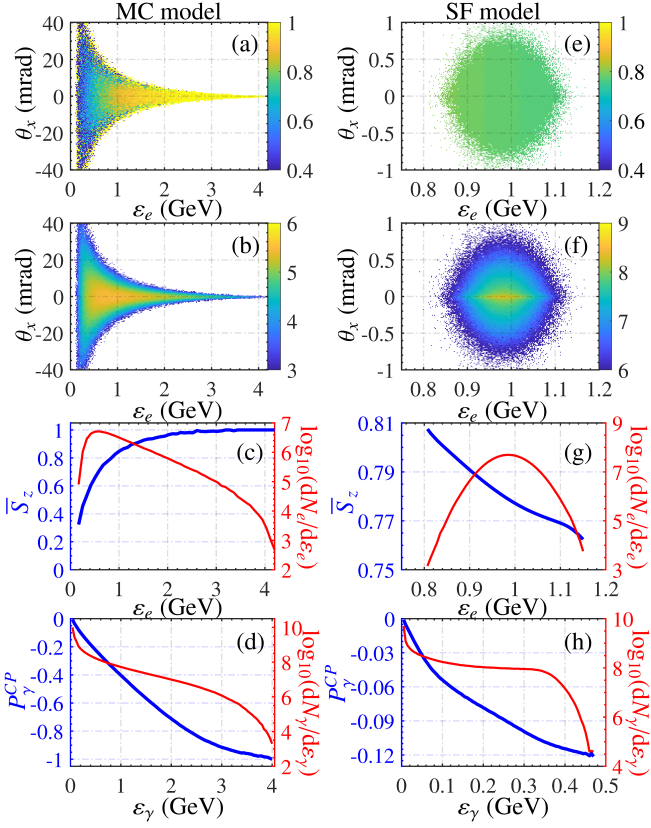}
	\caption{(a) and (e): Longitudinal  average spin (polarization) $\overline{S}_z$ vs the deflection angle $\theta_x$ =arctan($p_x/p_z$) and the electron energy $\varepsilon_e$. (b) and (f): Angle-resolved electron density log$_{10}$(d$^2N_e$/d$\theta_x$d$\varepsilon_e$) (mrad$^{-1}\cdot$ GeV$^{-1}$). (c) and (g): $\overline{S}_z$ (blue, calculated by summing over $\theta_x$ in (a) and (e), respectively) and log$_{10}$(d$N_e$/d$\varepsilon_e$) (red, calculated by summing over $\theta_x$ in (b) and (f), respectively) vs $\varepsilon_e$. (d) and (h): Degree of circular polarization of emitted  photons $P^{CP}_\gamma=\xi_2$ \cite{Ligammaray_2019, McMaster_1961} (blue) and energy density log$_{10}$(d$N_\gamma$/d$\varepsilon_\gamma$) (red) vs the  photon energy $\varepsilon_\gamma$. The left and right columns indicate the cases including and excluding radiative stochasticity, calculated by the MC and SF methods, respectively. The laser and electron beam parameters are given in the text.
	} \label{fig2}
\end{figure}

The angle- and energy-resolved  distributions of the polarization and density of the electron beam are illustrated
in Fig.~\ref{fig2}, including and excluding radiative stochasticity, calculated by the MC and SF methods, respectively.
Employed laser and electron beam parameters are as follows. A realistic tightly-focused Gaussian linearly polarized  laser pulse  \cite{Salamin2002, supplemental} propagates along $+z$ direction (polar angle $\theta_l=0^{\circ}$),   with   peak intensity  $I_0\approx3.45\times10^{21}$ W/cm$^2$ ($a_0=50$),  wavelength $\lambda_0=1$ $\mu$m,  pulse duration $\tau = 10T_0$ with  period $T_0$, and  focal radius $w_0=5$ $\mu$m. The counterpropagating LSP  electron beam has a cylindrical form,   with average spin (polarization) components $(\overline{S}_x, \overline{S}_y, \overline{S}_z) = (0, 0, 1)$, polar angle $\theta_e=180^{\circ}$, azimuthal angle $\phi_e=0^\circ$, radius  $w_e= \lambda_0$,  length $L_e = 5\lambda_0$, electron number $N_e=5\times10^6$ (density $n_e\approx 3.18\times10^{17}$ cm$^{-3}$ with
a transversely Gaussian and longitudinally uniform distribution), initial kinetic energy $\varepsilon_0=4$ GeV (the maximum value of the quantum parameter during the interaction is $\chi_{max}\approx 1.89$), angular divergence $ \Delta\theta = 0.3$ mrad, energy spread $\Delta \varepsilon_0/\varepsilon_0 =0.06$, and emittance $\epsilon_e\approx 3\times10^{-4}$  mm$\cdot$mrad.
 Such electron beams are achievable via laser wakefield acceleration \cite{Leemans2014,Leemans_2019} with further radiative polarization \cite{li2019prl,  Song_2019, Seipt_2019},
 or alternatively, via directly wakefield acceleration of LSP electrons \cite{Wen_2019,Wu_2019}.

Radiative stochasticity induces very broad angular and energy  distributions in the MC model in comparison with  the SF case, cf. panels ~(a) -(b) with  (e)-(f) in Fig.~\ref{fig2}. The spreads are particularly large in the laser polarization direction.  The  spin and density distributions of the electrons are demonstrated  more visibly for the MC model in Fig.~\ref{fig2}(c), by summing over $\theta_x$ in Figs.~\ref{fig2}(a) and (b), respectively. The electron energies after the interaction are distributed in the MC simulation in a rather large range  from 0.2 to 4.2 GeV, because due to the straggling effects some electrons do not radiate much.
 The average  spin polarization $\overline{S}_z$ monotonically increases with the energy from approximately 34\% up to 100\%. This is because  more photon emissions lead to more energy losses,
 and more radiation reaction to  more spin flips and further larger depolarization.

\begin{figure}[b]
\setlength{\abovecaptionskip}{-0.0cm}  	
	\includegraphics[width=1.0\linewidth]{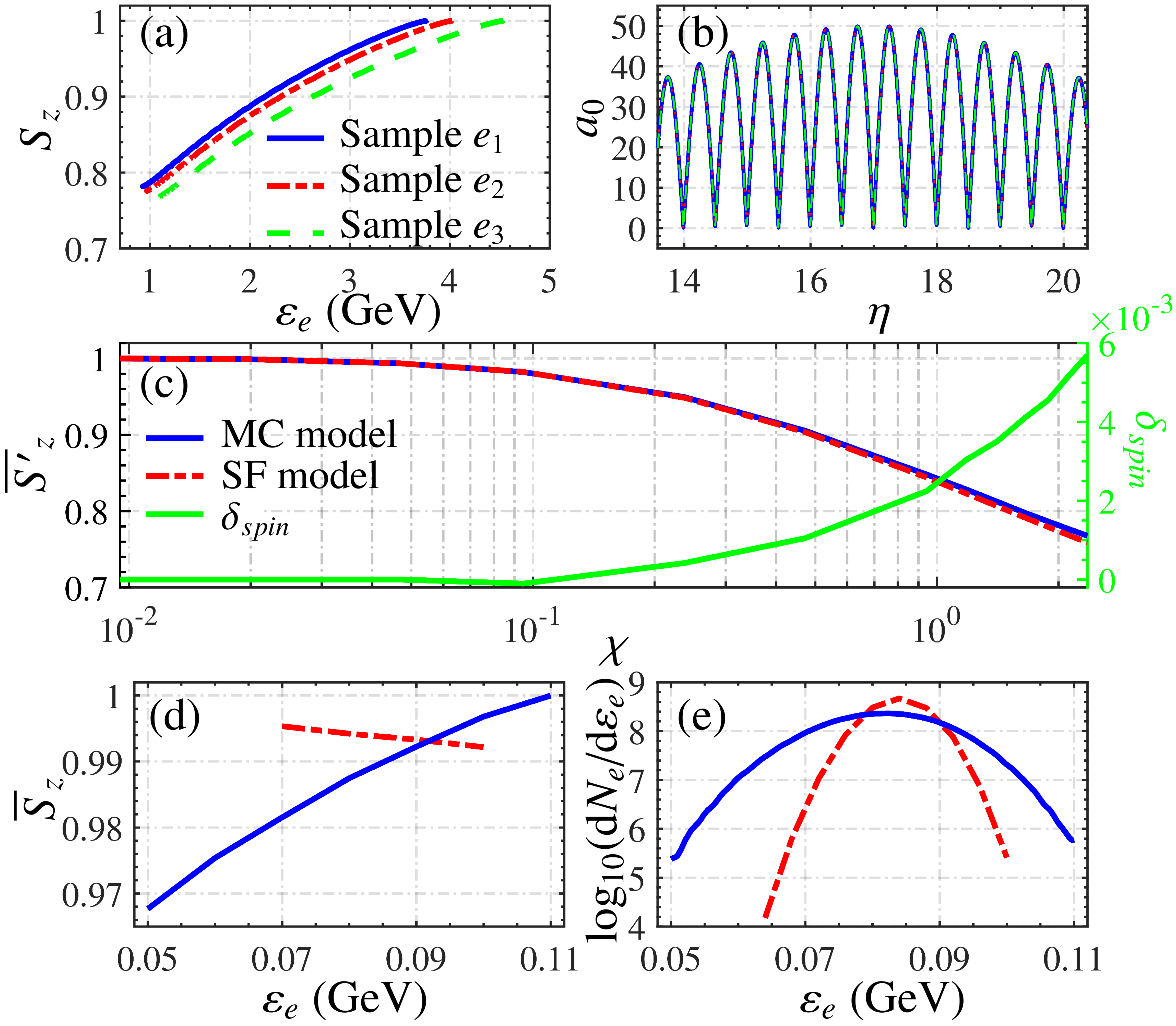}
		\caption{(a) and (b): Instantaneous $S_z$ vs $\varepsilon_e$ and experienced $a_0$ vs $\eta$ for three sample electrons, respectively, simulated by the SF method.  The sample electrons are chosen with randomly spatial coordinates and different energies. (c) The variation of the average polarization of all electrons $\overline{S'}_z$ and the relative deviation $\delta_{spin}$ with respect to  $\chi$. (d) and (e): $\overline{S}_z$ and log$_{10}$(d$N_e$/d$\varepsilon_e$) vs $\varepsilon_e$, respectively, for the case of $\chi\approx0.047$. In (c)-(e), the blue-solid and red-dashed curves are simulated by the MC and SF methods, respectively. Other laser and electron beam parameters are the same with those in Fig.~\ref{fig2}.
		}\label{fig3}
\end{figure}

In contrast to that, in the SF model the final electron energies have relatively small spread approximately from 0.81 GeV to  1.15 GeV. The $\overline{S}_z$  behaviour is qualitatively opposite to the MC model, it  monotonically decreases with the energy increase, but the variation is not large, approximately from
80.7\% to 76.4\%, as shown in Fig.~\ref{fig2}(g).
We analyze the reason for the polarization behaviour with the help of Fig.~\ref{fig3}.
First of all, let us note that the electrons in the beam experience similar instantaneous laser fields because the applied waist-size of the beam is not  small $w_0=5w_e$; see the fields experienced by three sample electrons in Fig.~\ref{fig3}(b). Then, the electron dynamics are gradually altered by continuous similar  photon emissions. The relation of the polarization to the energy during the interaction is shown in Fig.~\ref{fig3}(a).
For the electron  with a larger initial  energy (see  the sample electron ``$e_3$'' in Fig.~\ref{fig3}(a)), the radiation is stronger due to the larger parameter $\chi\sim a_0 \gamma_e$, and consequently, the depolarization is larger, but its final energy is still higher, because the  radiative energy loss is smaller than the initial energy spread.

The average polarization of all electrons $\overline{S'}_z$ in the MC and SF models are comparable, $\overline{S'}_z^{MC}\approx$  78.64\% and $\overline{S'}_z^{SF}\approx$  77.92\%, respectively, derived from data of Figs.~\ref{fig2}(c) and (g). The relative deviation  is $\delta_{spin}=(\overline{S'}_z^{MC}-\overline{S'}_z^{SF})/(\overline{S'}_z^{MC}+\overline{S'}_z^{SF})\approx0.46\%$.
The variation of  $\overline{S'}_z$ with respect to the quantum parameter $\chi$ is shown in Fig.~\ref{fig3}(c), which confirms that the SF method can provide the average depolarization (polarization) degree quite accurately, with a relative error of $\delta_{spin}<1\%$ at $\chi\lesssim 2$. With increasing $\chi$, the stochastisity effects become larger and  $\delta_{spin}$ raises. Although, at rather low $\chi\approx 0.047$, when the stochasticity is very weak, the average polarization can be deduced from the SF model, but the detailed  energy-resolved polarization and density still show differences with respect to the stochastic MC model, as shown in  Figs.~\ref{fig3}(d) and (e).
 Thus, the rising behavior of the electron polarization with the energy increase in the electron beam after the interaction [cf. panel (c) with (g) in Fig.~\ref{fig2}] is a distinct signature of the stochasticity in the radiative depolarization process.
Note that the polarization of employed  high-energy high-density electron beams can be detected by  the linear or nonlinear Compton scattering \cite{Narayan_2016,Li_2019spin}.

\begin{figure}[t]
	\setlength{\abovecaptionskip}{-0.0cm}  	
	\includegraphics[width=1.0\linewidth]{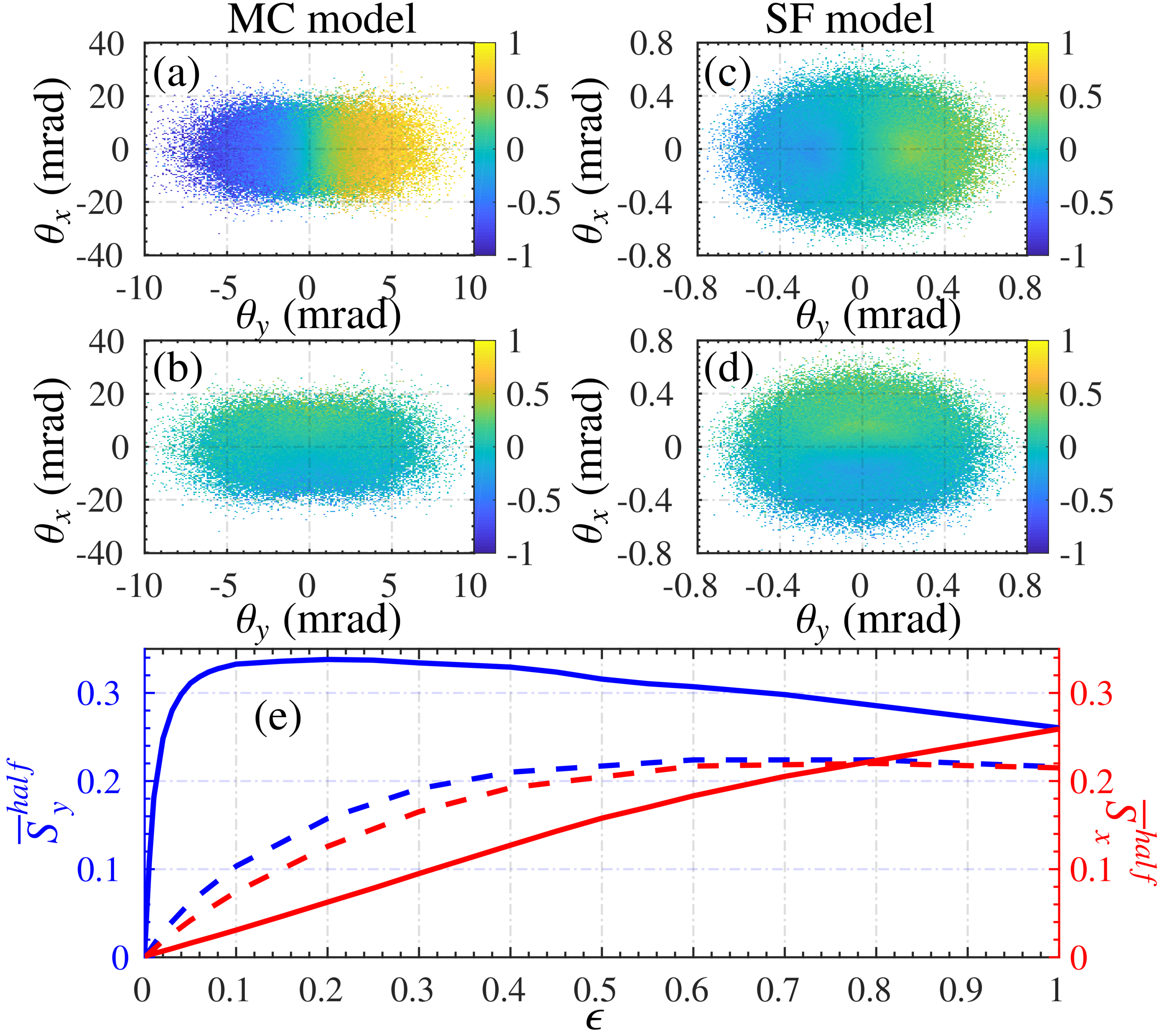}
	\caption{Transverse polarization of the initially unpolarized electron beam after the interaction with elliptically polarized laser field:(a) and (b) [(c) and (d)]: Transverse polarization components $\overline{S}_y$ and $\overline{S}_x$ vs the deflection angles $\theta_x$ and $\theta_y$ = arctan($p_y/p_z$), respectively, simulated by the MC [SF] method with the laser ellipticity $\epsilon=|E_y|/|E_x|$ = 0.2. 
(e) Average transverse  polarization $\overline{S}_y^{half}$ (calculated by summing $\overline{S}_y$  over  $\theta_y>0$ and $\theta_x$ in (a) and (c) for the MC (blue-solid) and SF (blue-dashed) methods, respectively) and $\overline{S}_x^{half}$   (calculated by summing $\overline{S}_x$ over  $\theta_x>0$ and $\theta_y$ in (b) and (d) for the MC (red-solid) and SF (red-dashed) methods, respectively) vs $\epsilon$.
Other laser and electron beam parameters are the same as in Fig.~\ref{fig2}.
	} \label{fig4}
\end{figure}

We have investigated also the role of stochasticity effects for emitted high-energy highly circularly-polarized $\gamma$-rays; see Figs.~\ref{fig2}(d) and (h).
While the circular polarization degree of $\gamma$-photons varies with  energy in a rather large range approximately from 0  to -1 in the MC model, the SF model shows much smaller range approximately from 0 to -0.1.  However, the average polarization degrees are similar and low, about -0.077 and -0.081 for the MC and SF models, respectively. This is because in the MC model the polarization is high for high-energy photons with very low numbers, see Figs.~\ref{fig2}(d)-(h).
 The energy range of $\gamma$-photons is much larger in the MC model, similar to the electron energy distribution, which yields  generation of high-energy high-brilliance highly circularly-polarized $\gamma$-rays, as discussed in \cite{Ligammaray_2019}.

Now we turn to the discussion of the case of initially unpolarized electron beam and look for the stochasticity effects in the spin dynamics. It is known \cite{li2019prl} that an initially unpolarized electron beam can be split into two oppositely polarized parts during interaction with a counterpropagating elliptically polarized laser pulse (the minor axis along $y$ direction).  We have analyzed this polarization-dependent splitting effect with the MC and SF models
for the full range of the ellipticity; see Fig.~\ref{fig4}. Exemplary distributions of the transverse polarization components with respect to the electron deflection angle after the interaction in  the case of $\epsilon=0.2$ are shown in Figs.~\ref{fig4}(a)-(d), calculated within the MC and SF models, respectively. In both models the electron beam splits into two parts along the propagation direction, which are oppositely polarized. At small ellipticity, the electron spin-polarization along the minor axis of the ellipticity is the largest, with small angular separation along that axis. The separated half of the electron beam (e.g. $\theta_y>0$) has an average polarization (e.g. $\overline{S}_y^{half}$), which depends on the separation angle: the larger  separation angle, the larger the average polarization $\overline{S}_y^{half}$. In the MC model the separation angle is significantly larger than that in the SF case due to stochasticity (as in this case photons of larger energies are emitted), and consequently $\overline{S}_y^{half}$ is larger; see Fig.~\ref{fig4}(e). The deviation of $\overline{S}_y^{half}$ between the MC and SF models  is the largest at small ellipticity near 0.05 for the given parameters. 
In the MC model of Fig.~\ref{fig4}(a) $|\overline{S}_y^{half}|\approx33.8\%$, by comparison in the SF model of Fig.~\ref{fig4}(c) $|\overline{S}_y^{half}|\approx15.8\%$ is much lower.
While in the SF model $\overline{S}_x^{half}$ and $\overline{S}_y^{half}$ increase monotonously with the increase of the ellipticity, in the MC model $\overline{S}_y^{half}$ demonstrates a characteristic nonmountainous behavior with a peak at small ellipticity. The latter can serve as a signature of the stochasticity effects in radiative polarization of  initially unpolarized electron beams.

For the experimental feasibility, we have investigated the impact of the laser and electron beam parameters, e.g.,  variations of $\varepsilon_0$, $a_0$ and $\tau$, larger energy spread of 10\%, larger angular divergence of 1 mrad, larger colliding angle of $\theta_e=175^\circ$ and initial transversely spin-polarization of the electron beam, on the considered signatures of radiative stochasticity, and the results keep uniform \cite{supplemental}.

In conclusion, we have analyzed the impact of stochastic photon emission in  a strong laser field on the initially LSP electron radiative depolarization as well as on the emitted $\gamma$-ray polarization. The qualitative signatures of the stochasticity have been demonstrated in the energy-resolved electron polarization after the interaction and in the energy-resolved polarization of the emitted $\gamma$-photons. In the case of initially unpolarized electron beam, the stochasticity effect is demonstrated in the dependence of the electron polarization on the laser ellipticity. These qualitative signatures are observable with the currently available laser facilities.\\

 {\it Acknowledgement:}  R.-T. Guo, Y. Wang and J.-X. Li thank Prof. C. Keitel for hospitality.
 This work is supported by the National Natural Science Foundation of China (Grants Nos. 11874295, 11875219 and 11905169), and the National Key R\&D Program of China (Grant No. 2018YFA0404801).

\bibliography{QEDspin}

\end{document}